\documentclass[9pt,twocolumn,twoside]{pnas-new}

\templatetype{pnasresearcharticle} 

\title{Learning-based link prediction analysis for Facebook100 network}

\author[a,b,1]{Tim Poštuvan}
\author[a,2]{Semir Salkić} 
\author[a,3]{Lovro Šubelj} 

\leadauthor{Poštuvan}

\affil[a]{University of Ljubljana, Faculty of Computer and Information Science, Ljubljana, Slovenia}
\affil[b]{University of Ljubljana, Faculty of Mathematics and Physics, Ljubljana, Slovenia}

\correspondingauthor{\textsuperscript{1} tim.postuvan@gmail.com \textsuperscript{2} ss9343@student.uni-lj.si \textsuperscript{3} lovro.subelj@fri.uni-lj.si}

\keywords{Link prediction $|$ Social networks $|$ Classification $|$ Supervised learning $|$ Feature selection} 

\begin{abstract}
In social network science, Facebook is one of the most interesting and widely used social networks and media platforms. Its data contributed to significant evolution of social network research and link prediction techniques, which are important tools in link mining and analysis. This paper gives the first comprehensive analysis of link prediction on the Facebook100 network. We study performance and evaluate multiple machine learning algorithms on different feature sets. To derive features we use network embeddings and topology-based techniques such as node2vec and vectors of similarity metrics. In addition, we also employ node-based features, which are available for Facebook100 network, but rarely found in other datasets. The adopted approaches are discussed and results are clearly presented. Lastly, we compare and review applied models, where overall performance and classification rates are presented.
\end{abstract}

\begin{document}

\maketitle
\thispagestyle{firststyle}
\ifthenelse{\boolean{shortarticle}}{\ifthenelse{\boolean{singlecolumn}}{\abscontentformatted}{\abscontent}}{}


\dropcap{S}ocial networks became an important focus in our research. We are witnessing exponential user expansion on social platforms (e.g. Facebook, Twitter and LinkedIn). People are joining these platforms and generating substantial amounts of data, which can reveal interesting clues about user behaviour, society and psychology. We can see significant increase in research on the topic of social networks and link prediction. In the last decade we are experiencing a rise in this overlapping topic of network science and data science, which is primarily used to analyze and understand social networks \cite{WangSurvey2014}. According to Facebook estimations in March 2020, they have 1.73 billion DAU (Daily Active Users). Much work has been done to understand complexities and challenges of social networks, where considerable knowledge has been obtained \cite{Pow2012, Soc1988}.
In that manner we are using the Facebook100 dataset (2005), which includes a complete set of people from Facebook networks for 100 different colleges and universities in the USA. This paper is the first comprehensive study of link prediction techniques applied to the Facebook100 dataset. 
We are using a learning-based approach with local probabilistic models such as logistic regression. Besides that we are utilizing standardized link prediction classification models such as: random forests, artificial neural networks and kernel based models (e.g. SVM). Having in mind that we have two main approaches to link prediction, we chose learning-based approaches instead of similarity-based approaches because we want to compare performance, stability and classification accuracy of multiple classification models instead of showcasing different measures of node proximity. We are using standardized similarity based methods to derive notable features, which are used as model input. The presented methods are node2vec graph embedding and an ensemble of topology-based metrics such as Jaccard Coefficient, Adamic Adar Coefficient and Preferential Attachment Index. Using these methods we build classification models which are clearly presented, focusing on performance and generalization, with appropriate discussion and results.

The paper is structured as follows. In section Related work we present related work in the field of link prediction on social networks. Section Data briefly describes the Facebook100 data. In sections Feature set and Datasets we present used feature sets and datasets, which are utilized for training and testing. Process and discussion of feature selection for each dataset are stated in the section Feature selection. Sections Results and Model analysis give an overview of achieved results, while also providing comprehensive model analysis. Insightful discussion of results which explains interesting nature of the data is explained in section Discussion. We conclude the paper with an overview in section Conclusion.

\section*{Related work}

Link prediction has recently become very popular for prediction of future relationships between individuals of social networks. Consequently, a great variety of different approaches were invented. In the past decade, many efforts have been made by psychologists, computer scientists, physicists and economists to solve the link prediction problem in social networks. According to Wang et al. \cite{WangSurvey2014} there are two ways to predict links: similarity-based approaches and learning-based approaches. Similarity-based approaches calculate a similarity score for every pair of nodes, where higher score means higher probability that the corresponding nodes will be connected in the future. 

Learning-based approaches are treating the link prediction problem as a binary classification task \cite{Hasan2006}. Therefore, typical machine learning models can be employed for solving the problem. These include classifiers like random forest \cite{RandomForest}, multilayer perceptron or support vector machine (SVM) \cite{SVM}, as well as probabilistic models. The learning-based approaches use non--connected pairs of nodes as instances with features describing nodes and the class label. Pairs of nodes which have potential to become connected are labeled as positive and the others as negative. 

Their feature set consists of similarity features from the similarity-based approaches and features derived from domain knowledge (e.g. textual information about members of social networks). Using combination of both can remarkably improve the link prediction performance. Scellato et al. \cite{Scellato2011} considered social features, place features and global features in location-based social networks for link prediction based on a supervised learning framework.

Both types of approaches rely on various metrics, which use information of nodes, topology of network and social theory to calculate similarity between a pair of nodes. Metrics consist of three categories: node-based, topology-based and social theory based metrics. 

Node-based metrics use the attributes and actions of individuals to assess similarity of node pairs. They are very useful in link prediction; however, it is usually hard to get the data because of privacy issues. 

Most metrics are based on the topological information and are called topology-based metrics. They are most commonly used for prediction, because they are generic and domain independent. Topology-based metrics are further divided into the following subcategories: neighbor-based, path-based and random walk based metrics. Neighbour based metrics assume that people tend to form new relationships with people that are closer to them. The most famous are Common Neighbors \cite{Newman2001}, Jaccard Coefficient \cite{Salton1986}, Adamic Adar Coefficient \cite{Adamic2003} and Preferential Attachment Index \cite{Barabasi2002}. The first three all use the same idea that two nodes are more likely to be connected if they share a lot of common neighbours. On the other hand Preferential Attachment Index assumes that nodes with higher degree have higher probability of forming new edges.

Neighbor-based metrics capture local neighbourhood but do not consider how nodes are reachable from one another. Path-based metrics incorporate this information by considering paths between nodes. They are more suitable to small networks and are not scalable to big networks. Examples of path-based metrics are Local Path \cite{Lu2009} and Katz metric \cite{Katz1953}. Local Path metric makes use of information of local paths with length two and three, while giving more importance to the paths of length two. Katz metric calculates the similarity by summing all the paths connecting the two nodes, giving higher weight to shorter paths.

Social interactions between members of social networks can also be modeled by random walk, which uses transition probabilities from a node to its neighbors to denote the destination of a random walker from the current node. Examples of random walk based metrics are Hitting Time and SimRank \cite{Jeh2002}. Hitting time metric calculates similarity based on the expected number of steps required for a random walk starting at a node to reach the other node. SimRank metric computes similarity according to the assumption that two nodes are alike if they are connected to structurally similar nodes. 

Social theory based metrics take advantage of classical social theories, such as community, triadic closure, strong and weak ties and homophily, improving performance by capturing additional social interaction information. Liu et al. \cite{Liu2013} proposed a link prediction model based on weak ties and degree, closeness and betweenness node centralities.

When designing a feature set, the choice of features tremendously influence the performance of link prediction. Sometimes is it hard to find appropriate features, hence it is desirable that an algorithm learns important features on its own. Network embedding methods aim at learning low-dimensional latent representation of nodes in a network. Embeddings should follow the principle that similar nodes in the network have similar embedding representations. The advantage of node embedding as a technique is enormous since it does not require feature engineering by domain experts. Network embeddings methods can be broadly categorized into four classes: methods based on random walks, matrix factorization, neural networks, and probabilistic approaches. For the purpose of this paper methods based on random walks are the most relevant. 

Methods based on random walks determine similarities using random walks on the original network. The Skip--Gram model, described in Mikolov et al. \cite{Mikolov2013}, is then usually used to generate node embeddings from the random walks. Examples of such methods are DeepWalk \cite{Perozzi2014} and node2vec \cite{Grover2016}. DeepWalk was the first technique for network embeddings, inspired by deep learning. It uses random walks with fixed transition probabilities to measure node similarity, while embeddings are derived using the Skip-Gram model. Node2vec is a generalization of DeepWalk which uses supervised random walks for node neighbourhood exploration. The random walk is controlled by a return parameter $p$ and an in-out parameter $q$. Then similarly Skip--Gram model is used, but this time approximated via negative sampling, for embedding generation. 

Evaluation of the methods plays the crucial role in machine learning task in general. To estimate performance of the link prediction approaches more evaluation criteria exists. While some papers utilize $Precision@N_p$ for a range of $N_p$ values \cite{Zhang2018}, others use AUROC \cite{Grover2016}.

\section*{Data}

We study the Facebook social network of friendships at one hundred American colleges and universities at a single moment of time in September 2005 \cite{facebook100, facebook5}. The network consist of one hundred independent networks, where every network corresponds to one university. Friendships are recorded only between people from the same university. Besides the information about friendships, network also contains limited demographic information. The following information is available for each user: student/faculty status flag, gender, major, second major/minor (if applicable), dorm/house, year and high school. Network is unweighted and undirected. The whole network consists of $3.2$ million nodes with $23.7$ million links between them \cite{facebook100statistics}. Maximum degree of a single node is approximately $4900$ and minimum degree is only $1$, with an average of $15$. According to statistics network appears to be disassortative but this is only the consequence of its size. It also has high average clustering coefficient $0.097$, which is characteristic of social networks.

\begin{table}[h!]
	\centering
	\begin{tabular}{ccccc}
		Dataset & $n$ & $m$ & $d$ &  $C$ \\
		\midrule
		Train  & 4943.5 & 206247.6 & 77.26787 & 0.2808 \\
		Unseen & 3517.8 & 140793.2 & 80.3072 & 0.2689 \\
		\bottomrule
	\end{tabular}
	\caption{Structural information of train and unseen data}
	\label{table:struct}
\end{table}

Table \ref{table:struct} contains structural measurements for used network sets. All of the presented measurements are averaged over all networks in the corresponding set. The presented values are: average clustering coefficient ($C$), average degree ($d$), average number of nodes ($n$) and average number of edges ($m$). We can see by the number of nodes and edges that train set is larger than unseen dataset. Clustering coefficient and average degree are high, which is one of expected characteristics of social networks.

\begin{figure}[h!]
	\centering
	\includegraphics[width=9cm, height=6.3cm]{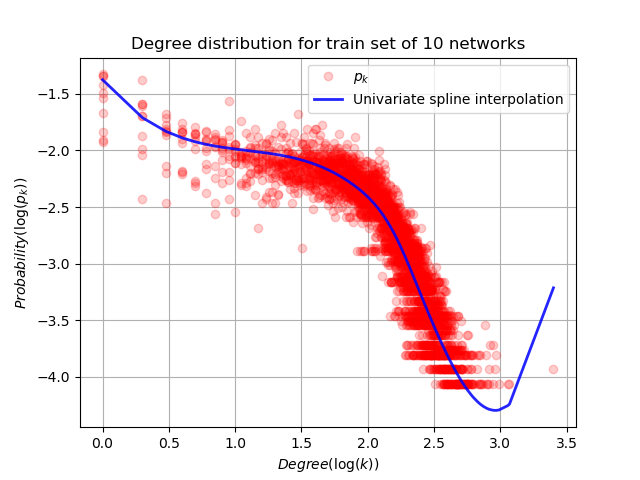}
	\caption{Degree distribution of train set}
	\label{fig:degtrain}
\end{figure}
\begin{figure}[h!]
	\centering
	\includegraphics[width=9cm, height=6.3cm]{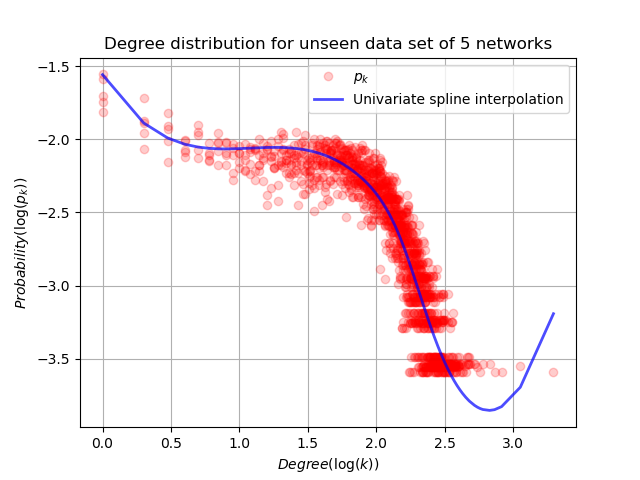}
	\caption{Degree distribution of unseen data set}
	\label{fig:degunseen}
\end{figure}

Because Facebook100 dataset is enormous, lack of computing power prevented us from considering the complete dataset for analysis. Therefore, we have decided that we would perform analysis only on a subset of networks. We selected ten networks as seen data and five networks as unseen data. Seen data was used for standard train and test set, where training and testing edges came from the same ten networks. On the other hand, unseen data consisted of five networks, which were not shown to the models during training. We used it to evaluate if our models are transferable to new data.

Degree distribution of the analyzed networks is presented in figures \ref{fig:degtrain} and \ref{fig:degunseen}. Degree distribution is plotted on a log-log scale for all networks in both sets respectively. For cleaner overview, we used interpolation (univariate spline) to showcase distribution of all networks. It is visible on both figures that all networks follow power law, which is expected for social networks. Having in mind that we are using real life social networks, it can be concluded that they are scale-free networks by degree distribution results. However, it is interesting to point out existence of big hubs, nodes with very high degree. They are visible on right side of the distribution graph. This is one of the reasons why interpolation at the end of the plot has an unexpected minimum.

\section*{Feature set}

Feature engineering probably plays the most important role when coping with a machine learning problem. Informative features crucially effect model accuracy, hence the process of feature engineering is usually very time consuming. In learning-based link prediction each pair of nodes is described using a combination of node-based, topology-based and node embedding features, depending on approach. In this paper we are using three different feature sets, from which three datasets were constructed as it will be described in the section Datasets.

\subsection*{Node-based features}

Node-based features use domain-specific information about individuals. Facebook100 dataset has already a basic set of features, however, not all of them are useful for link prediction task. Almost all features had to be transformed, in order to describe node pairs, instead of individuals. From some features, for example dormitory information, new features had to created, because otherwise model would not be transferable between networks. Problem arises from the fact that different universities use different numerations of their dormitories. Considering the above constraints, we derived the following features:

\begin{itemize}
	\item \textbf{same\textunderscore dorm}: binary value, indicating whether the nodes live in the same dormitory
	\item \textbf{same\textunderscore year}: binary value, indicating if the nodes started college in the same year
	\item \textbf{year\textunderscore diff}: numerical value, stating the absolute difference between the years, when the nodes started college 
	\item \textbf{high\textunderscore school\textunderscore 1, high\textunderscore school\textunderscore 2}: numerical values, stating indices of nodes' high schools 
	\item \textbf{major\textunderscore 1, major\textunderscore 2}: numerical values, stating indices of nodes' majors
	\item \textbf{same\textunderscore faculty}: binary value, indicating whether the nodes have the same faculty status
	\item \textbf{same\textunderscore gender}: binary value, indicating if the nodes have the same gender
\end{itemize}

Since networks are undirected, each pair of nodes must be uniquely represented using above features. Representation should not depend on order of the pair, thus major\textunderscore 1 and major\textunderscore 2 are ordered in a way that the value of major\textunderscore 1 is not greater than the value of major\textunderscore 2. The same holds for high\textunderscore school\textunderscore 1 and high\textunderscore school\textunderscore 2.

Like the majority of datasets Facebook100 does not contain all information about all individuals. Therefore, missing values had to be handled. We decided that imputing is reasonable only for the feature year\textunderscore diff, where missing values were substituted with the mean. Values of other features were left intact but as soon as one of the nodes in the pair had a missing value, the corresponding binary values was automatically zero.

\subsection*{Topology-based features}

The most commonly used features for link prediction are topology-based features. They are particularly useful, when you do not have any problem specific information, because they are generic and domain independent. Although Facebook100 dataset has additional domain specific data, topology-based features still have great impact on model accuracy. In this paper we are using the following topology-based features:

\begin{itemize}
	\item \textbf{Jaccard Coefficient} \cite{Salton1986}.
	Jaccard Coefficient normalizes the size of common neighbors. According to Jaccard Coefficient a pair of nodes is assigned a higher value when the nodes share a higher proportion of common neighbors relative to total number of their neighbours.
	
	\begin{equation}
	JC(x, y) = \frac{ |\Gamma(x) \cap \Gamma (y)| }{|\Gamma(x) \cup \Gamma (y)|}
	\end{equation}
	
	where $\Gamma(x)$ is a set of neighbours of node $x$.

	\item \textbf{Adamic Adar Coefficient} \cite{Adamic2003}.
	Adamic Adar Coefficient measure is closely related to Jaccard Coefficient. It is calculated as a weighted sum of common neighbours, where common neighbours with fewer neighbours have greater impact. The rationale behind it is that high degree nodes are more likely to occur in common neighbourhood, thus they should contribute less than low degree nodes. 
	
	\begin{equation}
	AA(x, y) = \sum_{z \in \Gamma(x) \cap \Gamma (y)} \frac{1}{\log |\Gamma(z)|}
	\end{equation}
	
	\item \textbf{Preferential Attachment Index} \cite{Barabasi2002}. The measure is based on the concept that nodes with higher degree have higher probability of forming new edges. 
	
	\begin{equation}
	PA(x, y) = |\Gamma(x)||\Gamma(y)|
	\end{equation}	
	
	\item \textbf{Resource Allocation Index} \cite{Zhou2009}.
	Resource Allocation Index metric is very similar to Adamic Adar Index. The only difference is that Resource Allocation Index punishes high degree nodes more.
	
	\begin{equation}
	RAI(x, y) = \sum_{z \in \Gamma(x) \cap \Gamma (y)} \frac{1}{|\Gamma(z)|}
	\end{equation}
	
\end{itemize}

\subsection*{Node embedding features}

Network embeddings methods aim to learn low-dimensional latent representation of the nodes in a network. To generate a dataset comprising of every node in a network we are able to use these representations as features. This can be used for a wide variety of tasks such as classification, clustering, link prediction, and visualization. Using node2vec \cite{Grover2016} we were able to generate our embeddings dataset.

The key point is that node2vec is based on random node walks performed in a biased manner across the network. With this generic approach we are able to sample any network in a search for vector representation of its structural properties. With the introduction of search bias $\alpha$ we are able to control our search in breadth-first search or depth-first search manner. If we choose \textit{“in-out parameter”} ($q$) , walks are more biased to visit nodes further from the start node, thus expressing the nature of exploration. Fixing \textit{“return parameter”} ($p=1$) ensures that we are less likely to visit same node twice, which in return adopts the strategy of modern exploration (avoids 2-hop redundancy in sampling).

As stated in the case study by Grover \& Leskovec \cite{Grover2016} for social structures it is beneficial to tune node2vec hyperparameters to discover communities of nodes which are interacting with each other. Capturing this type of behavior using embedding representation significantly benefits the link prediction task. To find the best set of hyperparameters, we employed grid search over more than 80 different settings that deemed reasonable to us. Each setting was evaluated on network of Caltech using logistic regression and the best combination of hyperparameters was selected. It is important to note here, that we did not consider only AUROC of the logistic regression model, but also the complexity of the embedding. The embedding dimension should be as small as possible while carrying all relevant information. The hyperparameters that were most consistent with the two criteria: 64 dimensions, 50 walks per node, $q = 0.8$ and 20 nodes in each walk. Since node2vec approach yields embeddings for nodes, we used Hadamard product to express vector representations for edges.

\section*{Datasets}

Firstly, we had to preprocess all graphs (seen as well as unseen) to obtain train and test node pairs, which will be predicted by models and using which performance will be evaluated. Node pairs can be either positive or negative instances for link prediction task, depending on whether there is an edge between the nodes or not (the nodes are friends or not). For every graph we used the standard approach of generating an incomplete train graph $G_{train} = (V, E_ {train})$ from the original graph $G = (V, E)$. The connected node pairs $\{i, j\} \in E \setminus E_{train}$, which are present in the original graph but not included in the train graph, are used as positive instances for link prediction task. Positive instances were sampled randomly from the original graph's edge set $E$. We decided to sample 2\% of edges in original graph $G$. Since dataset should contain positive as well as negative instances, we had to obtain also negative instances -- pairs of nodes that are not connected by an edge. It is assumed that if two nodes are not connected by an edge, they are not friends. Negative instances were obtained by randomly selecting node pairs $\{i, j\} \notin E$, which are not in the original graph's edge set. To get a balanced dataset the number of negative instances is the same as the number of positive ones. The rest of the graph ($G_{train}$) was used to calculate topology-based and node embedding features. Positive train and test edges were not included in this graph from which features were derived, otherwise we would introduce unjust label information in features. 

In our experiments, all unseen data instances were used for testing models' ability to adapt to new graphs. However, seen data was further split into train and test data. We used standard division: 80\% of it was used as the train data and the remaining 20\% was used as the test data. Within each of the them there is approximately the same number of positive and negative instances.

Using this data three datasets were created: baseline, topological and embedding dataset. Each dataset represents node pairs using a different combination of features. Baseline dataset is the simplest one and contains only topology-based features. A bit more complex is topological dataset, which in addition to the topology-based features makes use of node-based features as well. Node pairs in embedding dataset are described using node-based features and Hadamard product of the corresponding nodes' embeddings.

\begin{SCfigure*}[\sidecaptionrelwidth][t]
\centering
\includegraphics[width=16cm]{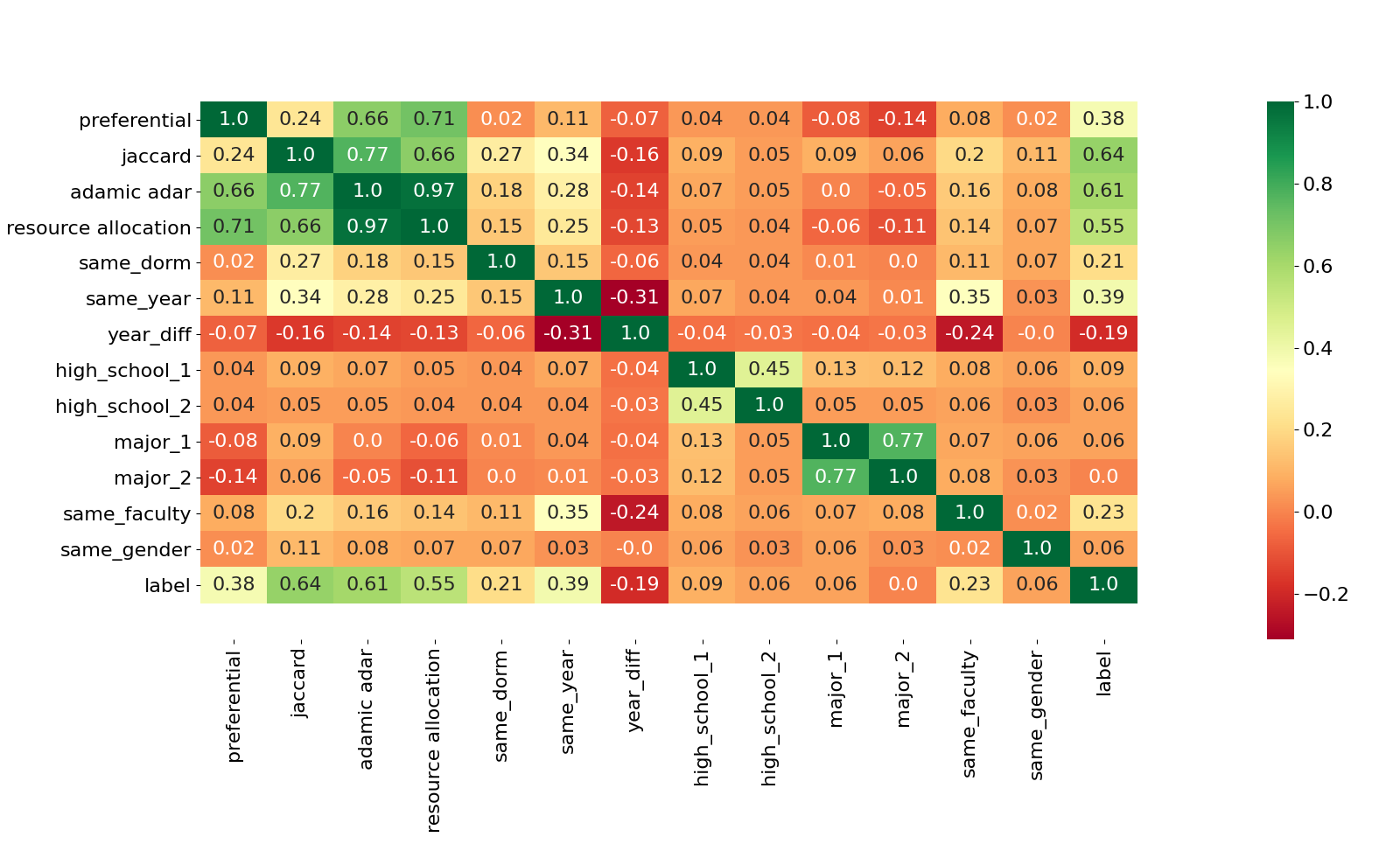}
\caption{Correlation matrix of node-based and topology-based features}
\label{fig:correlation_matrix}
\end{SCfigure*}

\section*{Feature selection}

Contemporary datasets usually have abundance of data, which is not always relevant to the problem. Hence, datasets should be preprocessed before models are used on them. Preprocessing takes place mainly to reduce the size of the dataset and achieve more efficient analysis, as well as removing redundant features, which have negative impact on the performance of the model. The aim of feature selection is to maximize relevance and minimize redundancy of the features. 

Our feature sets are not enormous, thus feature selection was done solely for the sake of performance improvement of the models. We are using recursive feature elimination with cross-validated selection (RFECV) in combination with linear kernel support vector machine (SVM) to get reduced feature sets. This method recursively considers smaller and smaller sets of features, while after every iteration prunes the least important features according to the chosen model. It belongs to wrapper methods for feature selection, since it appraises subsets of features based on performance of the modelling algorithm. According to Jović et al. \cite{Jovic2015} wrapper methods have been empirically proven to yield better results than other methods because subsets are evaluated using real models.

\subsection*{Baseline dataset}

\begin{figure}[H]
	\centering
	\includegraphics[width=9cm, height=7cm]{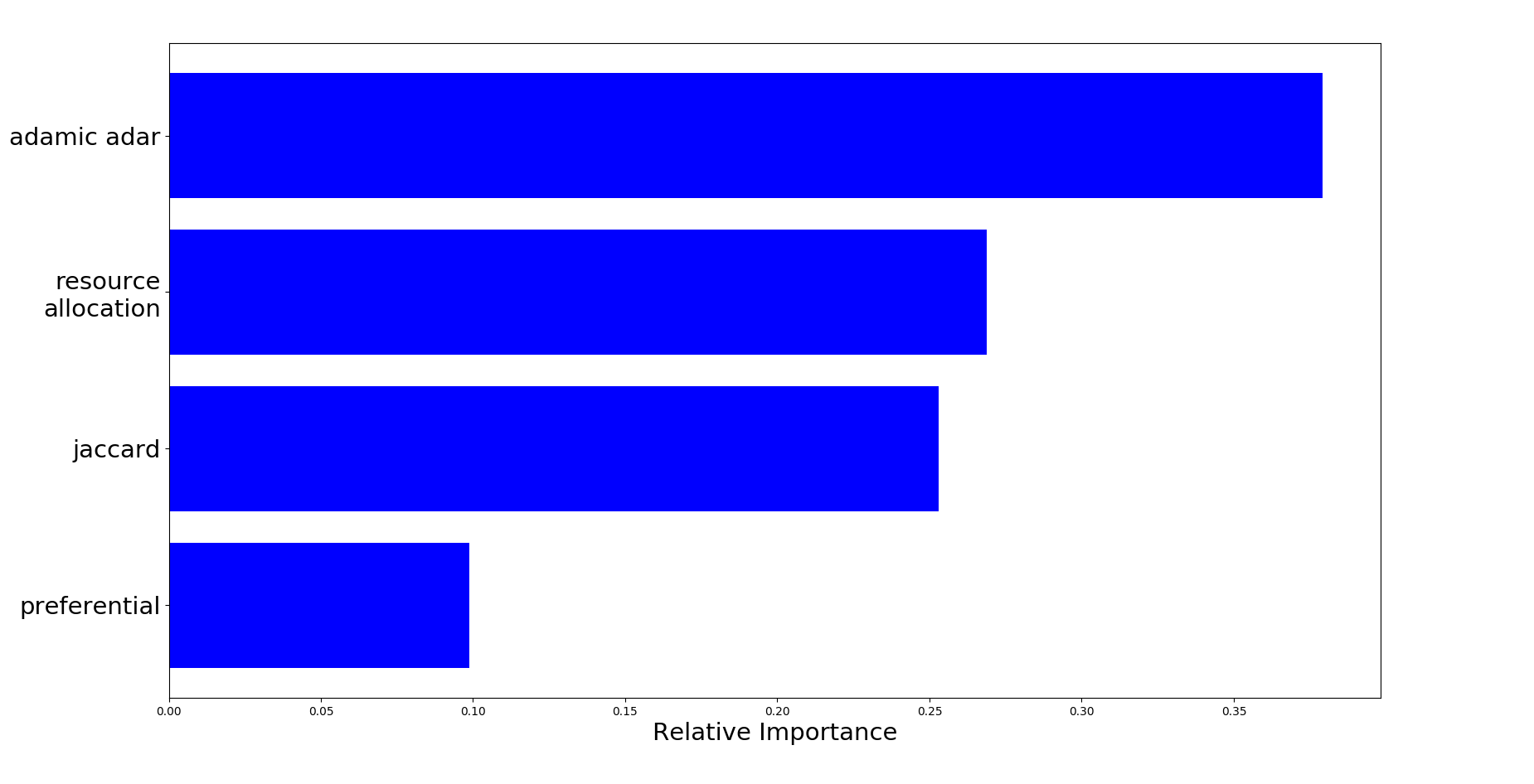}
	\caption{Feature importance of baseline dataset according to random forest classifier}
	\label{fig:baseline_feature_importance}
\end{figure}

The above feature selection method recognized Adamic Adar Coefficient, Jaccard Coefficient and Resource Allocation Index as the most informative features. The most relevant feature is Adamic Adar Coefficient and the least relevant one is Preferential Attachment Index. This is completely coherent with random forest feature importance shown in figure \ref{fig:baseline_feature_importance}. Adamic Adar Coefficient is the most relevant feature, although Jaccard Coefficient has higher correlation with labels, which can be observed in figure \ref{fig:correlation_matrix} . All selected features are highly correlated with label, whereas Preferential Attachment Index is not. This is probably the reason why Preferential Attachment Index is the only feature which was not selected. From correlation matrix it is also evident that Adamic Adar Coefficient and Resource Allocation Index are almost perfectly correlated, which is expected because of the similarity in their definitions. Nonetheless, adding it results in a slightly better performance, thus the algorithm decides to keep it.

\subsection*{Topological dataset}

\begin{figure}
\centering
\includegraphics[width=9cm, height=7cm]{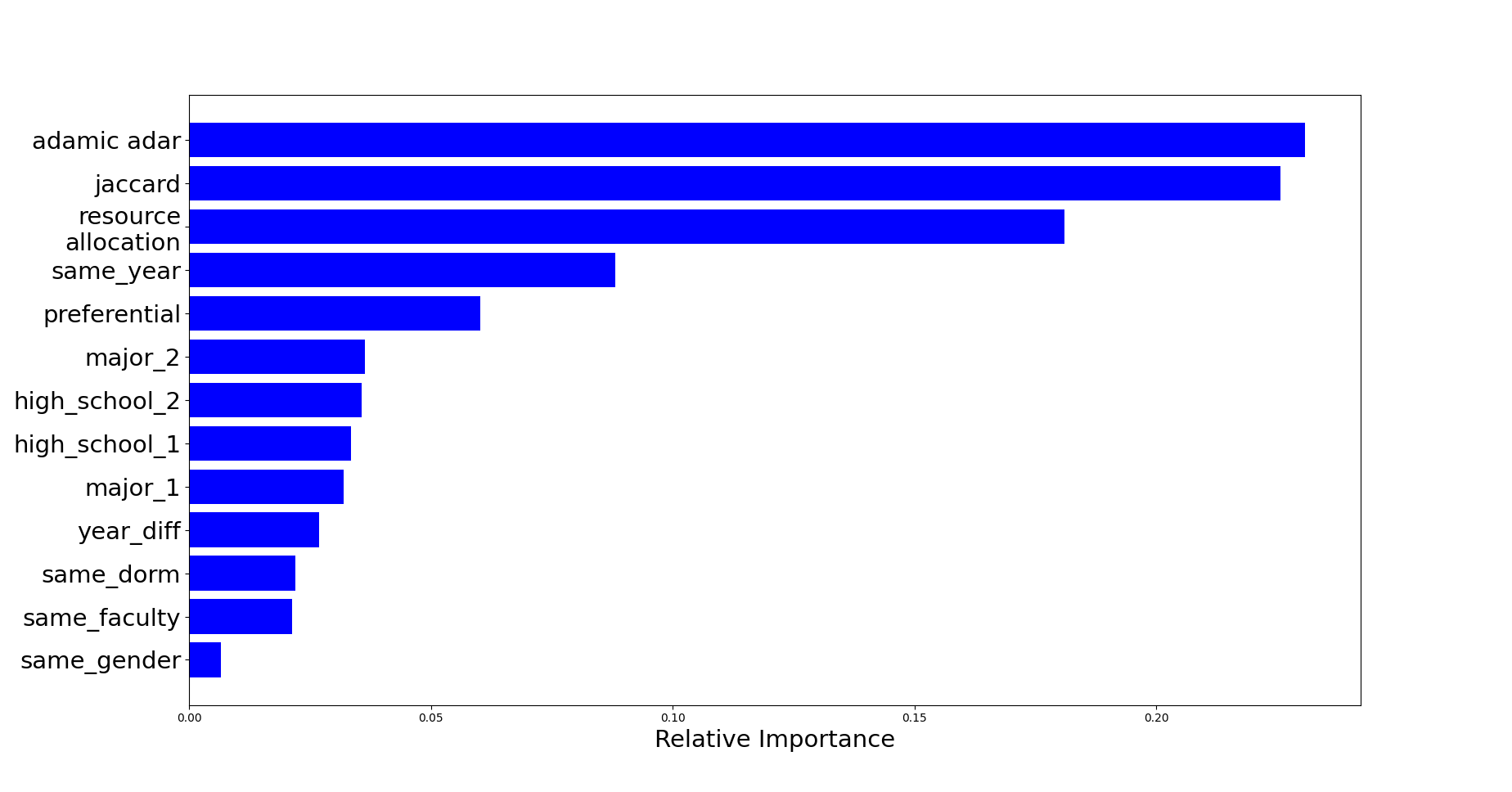}
\caption{Feature importance of topological dataset according to random forest classifier}
\label{fig:topological_feature_importance}
\end{figure}

The advantages of feature selection are more evident on topological dataset, because it has more features. This time algorithm selected the following features: all four topology-based features, same\textunderscore year, same\textunderscore faculty, same\textunderscore dorm, major\textunderscore 1 and major\textunderscore 2. On figure \ref{fig:topological_feature_importance} it is clearly shown that topology-based features are far more important than other node-based features. This is also consistent with correlation matrix, since topology-based features have the highest correlations with labels. They are so informative due to phenomenon called triadic closure. The triadic closure states that in social networks connections tend to form between people who share common friends, which is precisely what these topology-based features are describing. Among the node-based features same\textunderscore year, same\textunderscore faculty and same\textunderscore dorm were selected, all having relatively high correlation with label. Particularly high correlation has same\textunderscore year, which is expected, as college students often form friendships with their classmates. Because of this major\textunderscore 1 and major\textunderscore 2 are also relevant. Feature same\textunderscore faculty exploits the fact that students' and professors' social circles are rarely overlapping.

\subsection*{Embedding dataset}

Feature selection on embedding dataset was especially hard due to artificial features from node2vec. Because hyperparameters of node2vec were carefully tuned, we assumed that node embeddings are optimal. We selected the minimum embedding size which still performed well, while the quality of a set of hyperparameters was evaluated on the link prediction task. This was the best we could do with available computational resources, since there were too many features to utilize the feature selection method described above. Hence, we filtered only node-based features \cite{Attributed} and are using only a few crucial ones: same\textunderscore year and same\textunderscore dorm. We did not select same\textunderscore faculty, although it is more important than same\textunderscore dorm if considered on its own. We decided so, because same\textunderscore faculty has high correlation with same\textunderscore year and correlated features usually have negative impact on performance of the model.

\section*{Results}

Evaluation of our datasets was conducted using an ensemble of classification models. We used simpler models like logistic regression and random forest, as well as more complex ones -- support vector machines (SVM) and neural networks (NN), which are capable of modeling more complex non-linear functions. Hyperparameters of all models were optimized as described in section Model analysis. Link prediction task was tested on all three datasets, on test and unseen data, and all aforementioned models. Performance of the models was evaluated using Area Under the Receiver Operating Characteristics (AUROC), which is one of the most common evaluation metrics for link prediction.

\begin{table}[h]
	\centering
	\begin{tabular}{ccccc}
		Dataset & Logistic regression & Random forest & SVM & NN \\
		\midrule
		Baseline & 0.9401 & 0.9227 & \textbf{0.9628} & \textbf{0.9618} \\
		Topological & 0.9570 & 0.9173 & \textbf{0.9639} & \textbf{0.9623} \\
		Embedding & 0.9365 & 0.9145 & \textbf{0.9414} & 0.9389 \\
		\bottomrule
	\end{tabular}
	\caption{AUROC values for logistic regression, random forest, support vector machine (SVM) and neural network (NN) on test data}
	\label{table:normal_results}
\end{table}

\begin{table}[h]
	\centering
	\begin{tabular}{ccccc}
		Dataset & Logistic regression & Random forest & SVM & NN \\
		\midrule
		Baseline & 0.9263 & 0.9031 & \textbf{0.9570} & \textbf{0.9560} \\
		Topological & 0.9478 & 0.8901 & \textbf{0.9563} & 0.9538 \\
		Embedding & \textbf{0.9229} & 0.9047 & \textbf{0.9217} & \textbf{0.9218} \\
		\bottomrule
	\end{tabular}
	\caption{AUROC values for logistic regression, random forest, support vector machine (SVM) and neural network (NN) on unseen data}
	\label{table:unseen_results}
\end{table}

Table \ref{table:normal_results} contains AUROC scores for all combinations of the datasets and models on the test data. Similarly, table \ref{table:unseen_results} states the same values, but on unseen data. These tables reveal that support vector machine (SVM) and neural network (NN) are the best models for the link prediction task. Their performance is almost exactly the same, although they are based on completely different concepts. This is indicating that all relevant information from datasets is used for prediction. Only a little worse did logistic regression, which is very surprising, since it is much simpler than SVM and NN. Even more unexpected is that it outperformed random forest, which is a non-linear model. We hypothesize that this is a consequence of linear separability of the data, but more about this will be written in section Discussion. All models appear to be stable, since there is only a slight decrease in performance, when applied to unseen data. Difference is negligible for baseline and topological dataset, whereas noteworthy on embedding dataset. 

The models were able to extract more useful information from topological dataset than baseline and embedding ones. Baseline dataset has only a bit worse results, showing that additional node-based features have minimal influence on performance. Difference is visible for logistic regression, whereas SVM and NN have the same score on both datasets. Shockingly, embedding dataset gets the worst results. However, this might be the consequence of the chosen evaluation metric. For example logistic regression on embedding dataset gets $0.87$ $F_1$ score, while baseline and topological datasets get only $0.76$ and $0.79$. $F_1$ score reflects balance between recall and precision, whereas AUROC does not consider the predicted labels, but only measures whether predicted values of negative instances are smaller than predicted values of positive instances. Consequently, metrics are not necessary coherent, however, we decided to trust AUROC since it is a more standard link prediction metric.

\section*{Model analysis}

Here we used previously obtained data and results to optimize our models. Prior to that, data was standardized to have variance $1.0$ and mean $0.0$. With this approach, we have done model analysis to interpret best combinations of hyperparameters, which are useful to understand and discover patterns in data.

\subsection*{Logistic regression}

Using grid search cross-validation on logistic regression we saw that different approaches require different configurations. In the case of the baseline approach features equally impact the decision process, which is reflected in features' coefficients. For this dataset we used ridge regularization ($L_2$) with default regularization strength of $1$.

In the case of topological dataset we noticed that regularization significantly influences performance. We used lasso regularization ($L_1$) regularization with immense regularization strength of $1000$. In this case regularization is crucial for prevention of overfitting. Having in mind that our features are measurements which are not calculated in a fixed interval, we are benefiting from the $L_1$ property of data sparsity. Model coefficients are imbalanced, where major study features (i.e. major\textunderscore 1 and major\textunderscore 2) are given low coefficient values, which shows that most of the information is contained in the rest of the features.

In embedding dataset we noticed that addition of node-based data does not have any benefits. This could be the consequence of correlations within features of embedding dataset. Maybe node2vec features contain structural information, using which node-based feature can be well approximated. For embedding dataset $L_1$ regularization with strength $150$ was used. Lasso regularization improves model performance on unseen networks. This is achieved with generalization of the obtained knowledge from social network onto new unseen networks. As expected, coefficients show that all features of embedding vectors are equally important.

\subsection*{Random forest}

Random forest did not perform well on our datasets. We can justify that by the fact that this approach lacks mechanism for regularization. Higher number of dimensions in respect to number of samples (unbalanced training and unseen data) is causing our decision tree models to overfit. Grid search in this case did not yield specific results, as well as tuning of parameters failed to find feature dependent information. This behaviour is shown in our model comparison where it is expected to experience better benchmarks on different linear models such as logistic regression and SVM. We noticed that unseen networks' AUROC scores are the lowest over all datasets, therefore we can conclude that random forest model did not respond well to our problem.

\subsection*{Support vector machine (SVM)}

For support vector machine (SVM) only kernels were carefully tuned. Best fit for each dataset was chosen using grid search. Grid search consisted of linear, polynomial and Gaussian kernel, so the model could work with arbitrary dimensional data. It turned out that for baseline and topological datasets linear kernel was the best option, while embedding dataset required Gaussian kernel.

\subsection*{Neural network (NN)}

Choosing the right hyperparameters for neural networks (NN) was very complicated and tedious task, since neural networks have a lot of different parameters. Nevertheless, correctly setting them can yield much better performance in comparison to other models. For some of the parameters like loss and optimization functions default settings were selected. Because learning-based link prediction is a binary classification task, binary cross-entropy loss function and Adam optimizer were utilized. Hidden and output activation functions were selected using random experimentation. The best results yielded ReLU as hidden activation function and sigmoid as output activation function. Lastly, architecture of neural network had to be defined, which was done using grid search. We tried a great variety of different depths and numbers of nodes per layer, but in the end architectures with only two hidden layer and small number of nodes were the best performing on all datasets.

\section*{Discussion}

Embedding dataset yielded worse results than topological and baseline ones. This could be so because of greater linear separability of topological and baseline datasets as SVM results from section Results were implying. It is much easier to train models on linearly separable data than on complex one with a lot of non-linearity, such as node2vec.

\begin{figure}[h!]
	\centering
	\includegraphics[width=9cm, height=6.5cm]{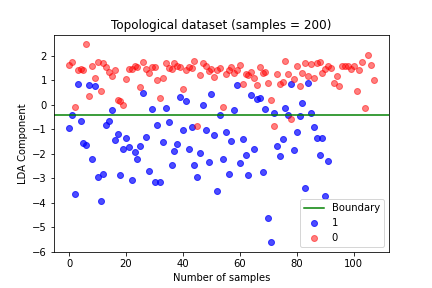}
	\caption{LDA on topological dataset}
	\label{fig:toplda}
\end{figure}

Baseline and topological data seem to be well linearly separable, because SVM works best on them when combined with linear kernel. Also, logistic regression performs considerably better than random forest. This is highly unusual for non-linear datasets, but in this case random forest harder adjusts to linear data, while logistic regression is linear by default.

Trying to prove our hypothesis that topological dataset is more linearly separable than embedding one, we visualized data using linear discriminant analysis (LDA). If any linearity exists in the dataset, it should be visible in the low-dimensional space. To visualize data we uniformly sampled $200$ samples from test data of both datasets and calculated LDA decision boundary for binary classification. 

\begin{figure}[h]
	\centering
	\includegraphics[width=9cm, height=6.5cm]{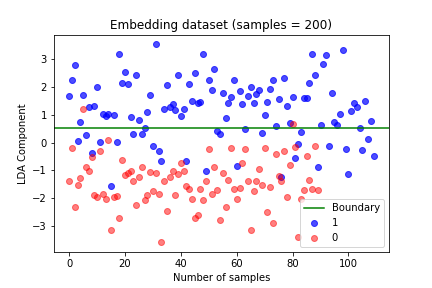}
	\caption{LDA on embedding dataset}
	\label{fig:emblda}
\end{figure}

LDA analysis seems to show that the data is approximately equally linearly separable in topological dataset (figure \ref{fig:toplda}) as well as in embedding one (figure \ref{fig:emblda}). Although, accuracy score of primitive LDA classifier yielded slightly better results on topological test set $87\%$, in comparison to embedding one which had score of $86.5\%$. A more concrete evidence of worse linear separability is probably the fact that SVM with Gaussian kernel performed better on embedding dataset than SVM with the linear kernel. 

Nonetheless, it appears that embedding dataset is still linearly separable to some degree. This is completely unexpected as node2vec produces inherently non-linear embeddings. These interesting results pose a new research question that could be examined in future work.

Besides assumed linearity, topological dataset could also give better results due to small number of features ($9$ features), whereas embedding dataset was significantly more complex ($66$ features). It is harder to train models on data with a lot of features, because models automatically require more parameters, so worse performance of embedding dataset could also stem from that.

\section*{Conclusion}

In the presented paper we conducted the first comprehensive analysis of link prediction on the Facebook100 network. We can conclude that results are unexpectedly good for link prediction task of this nature. After all, we are predicting friendships on completely separated social networks. It is visible that models successfully generalized to unseen data, even though train set is only 50\% bigger than unseen data set. Therefore, it is safe to say that our models succeeded in their task.

For optimization of AUROC score baseline and topological approaches are the best. It turns out that simplicity has benefits in terms of higher classification scores. In these two cases node-based features did not really effect performance, except when combined with logistic regression. Although SVM and NN got better results, we recommend logistic regression in combination with topological approach because the model is easier to train and interpret. When very high AUROC scores are important (e.g. link prediction on medical data), we suggest SVM with linear kernel and baseline approach. It gets almost the same results on unseen data, even though it is simpler model than NN.

We have shown that collecting data from multiple social networks yields promising datasets, which can be used for modelling of various predictors in similar social structures. Besides that, in this paper we have shown that use of regularization can be a solution in the case of social networks, when lack of the training data is present. Using this approach we can obtain data insights globally.

For future work it would be interesting to show how much linearly separable is embedding dataset. Even more fascinating would be to find the true underlying cause of the observed behaviour.

\section*{References}

\bibliography{link_prediction}

\end{document}